\documentclass[prb,twocolumn,showpacs,preprintnumbers,amsmath,amssymb,floatfix]{revtex4-1}

\usepackage{graphicx,color}

\begin{document}

\title{Band structure and optical transitions in atomic layers of hexagonal gallium chalcogenides}

\author{V.\ Z\'{o}lyomi, N.\ D.\ Drummond, and V.\ I.\ Fal'ko}

\affiliation{Physics Department, Lancaster University, Lancaster LA1 4YB,
  United Kingdom}

\date{\today}

\begin{abstract} We report density-functional-theory calculations of the
electronic band structures and optical absorption spectra of two-dimensional
crystals of Ga$_2$X$_2$ (X=S, Se, and Te).
Our calculations show that all three two-dimensional materials are dynamically stable
indirect-band-gap semiconductors with a Mexican-hat dispersion of holes near
the top of the valence band. We predict the existence of Lifshitz transitions---changes
in the Fermi-surface topology of hole-doped Ga$_2$X$_2$---at hole
concentrations $n_{S}=7.96\times 10^{13}$ cm$^{-2}$, $n_{Se}=6.13\times 10^{13}$ cm$^{-2}$,
and $n_{Te}=3.54\times 10^{13}$ cm$^{-2}$.

\end{abstract}


\maketitle

\section{Introduction}

Two-dimensional semiconductors have been studied intensively in
the past few decades as researchers have tried to find  materials to
complement or even replace silicon in electronics.
A recent trend in the hunt for new two-dimensional systems consists of the
isolation and study of atomically thin sheets of layered materials.
Honeycomb carbon allotropes have been widely investigated for this purpose following the discovery of
graphene \cite{novoselov_2004,geim_2007}; hexagonal boron nitride is another
example \cite{BN01,BN02}.
With increasing chemical unit cell complexity, hexagonal
transition metal dichalcogenides have been exfoliated into atomically thin films
using mechanical transfer and liquid-phase sonication \cite{mos2exfol,KisA_2011_2,KisA_2011,WS2exp01,WS2exp02};
many of these materials are semiconductors in both bulk and monolayer phases, making them ideal for use in
single-sheet nanoscale optoelectronics \cite{KisA_2011_2,AtacaC_2012,WS2exp01,Morpurgo}. In this
work we explore the next generation in the family of two-dimensional atomic crystals:
stoichiometric layers of gallium chalcogenides (Ga$_2$X$_2$).

\begin{figure}
\begin{center}
\includegraphics[clip,scale=0.2,angle=0]{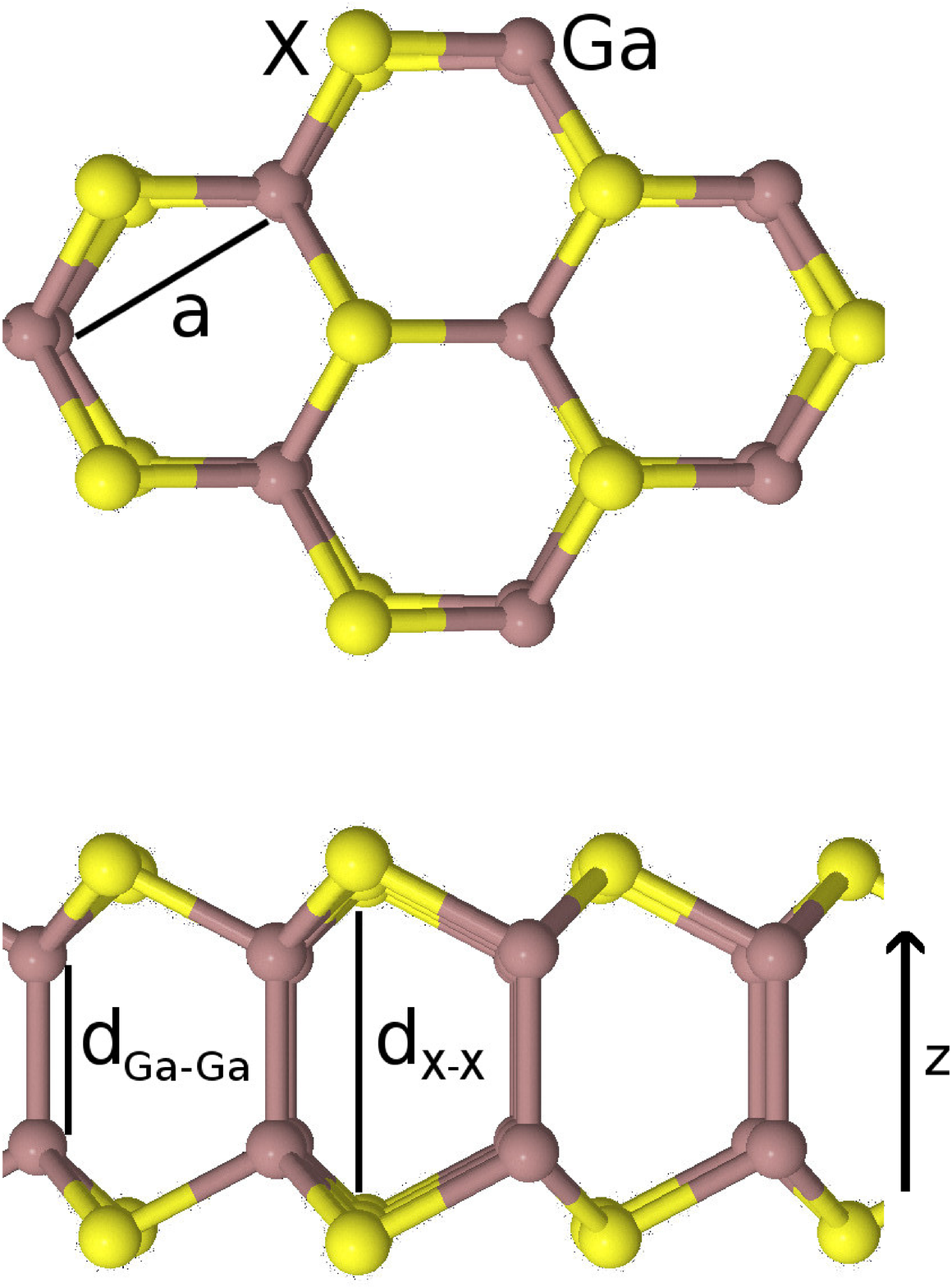}
\caption{(Color online) Structure of monolayer gallium chalcogenides Ga$_2$X$_2$ (X=S, Se, Te) in top and side views.
\label{fig:structure}
}
\end{center}
\end{figure}

In the bulk form, GaS and GaSe are indirect-gap semiconductors
\cite{ho_ch_2006}, with the latter
being well-known for its nonlinear optical properties \cite{bhar_gc_1995},
while GaTe is a direct-gap semiconductor used for electrothermal
threshold switching \cite{Milne_WI_1973}. All three
materials exhibit the same structure: they are layered compounds of Ga$_2$X$_2$ stoichiometry
in which each layer
consists of two AA-stacked hexagonal sublayers of
gallium atoms sandwiched between two hexagonal sublayers of chalcogen atoms (X), as illustrated
in Fig.\ \ref{fig:structure}. In a bulk material, these
layers are bound in a three-dimensional structure by van der Waals interactions.

In this work we discuss the optical and electronic properties of single-layer crystals of
Ga$_2$S$_2$, Ga$_2$Se$_2$, and Ga$_2$Te$_2$. We present the electronic
band structures and optical absorption spectra for all three materials. In
each case we find that the band gap is indirect,
with the valence-band maximum shifted from the $\Gamma$ point
(an inverted Mexican-hat dispersion) and
the conduction-band minimum (CBM) located at the M point in Ga$_2$S$_2$ and Ga$_2$Te$_2$, and at the
$\Gamma$ point in Ga$_2$Se$_2$. For all three materials we find an unusual, slightly anisotropic
Mexican-hat shape in the electron dispersion near the edge of the valence band.
However, the strongest peak in the optical absorption spectra---dominated
by a transition between two bands with states which are even with respect to $z\rightarrow -z$ symmetry---is
not caused by the Van Hove singularity associated with the Mexican-hat spectra,
but is provided by two, near-parallel branches in the dispersion of the conduction bands
near the Brillouin-zone corners K and K$^\prime$. We also find that hole-doped Ga$_2$X$_2$ 
undergoes a Lifshitz transition\cite{Lifshitz1960}---a change in the topology of the Fermi surface---at hole
concentrations $n_{S}=7.96\cdot 10^{13}$ cm$^{-2}$, $n_{Se}=6.13\cdot 10^{13}$ cm$^{-2}$,
and $n_{Te}=3.54\cdot 10^{13}$ cm$^{-2}$.

\section{Results and discussion}

The band-structure analysis of single-layer Ga$_2$X$_2$ leading to the above conclusions was
performed using density-functional-theory (DFT) as implemented in
the \textsc{castep} \cite{castep} and \textsc{vasp} \cite{vasp} plane-wave-basis
codes. Tests have shown that the two codes yield nearly indistinguishable results
for the materials studied here.
To calculate the geometries, phonon dispersions, and optical absorption spectra,
we used semilocal exchange-correlation functionals: the local density approximation (LDA)
and the Perdew-Burke-Ernzerhof \cite{pbe} (PBE)
functionals. The screened Heyd-Scuseria-Ernzerhof 06 (HSE06) functional
\cite{hse} was used to obtain the electronic band structures
to compensate at least partially
for the underestimation of the band gap by semilocal functionals.
The plane-wave cutoff energy was 600 eV\@. During relaxations a $12\times 12$ Monkhorst-Pack
\textbf{k}-point grid was used, while band structures were obtained with a $24\times 24$ grid. The
optical absorption spectra were obtained with a very dense grid of $95\times 95$ \textbf{k}-points.
The artificial out-of-plane periodicity of the monolayer was set to 20~\AA\@.
Phonon dispersions were calculated in \textsc{castep} \cite{castep2} using the method of finite displacements in
a $4\times 4$ supercell, with $11\times 11$ \textbf{k}-points,
a 408 eV plane-wave cutoff, and an artificial out-of-plane periodicity of 15.88~\AA\@.

\begin{figure}
\begin{center}
\includegraphics[clip,scale=0.8]{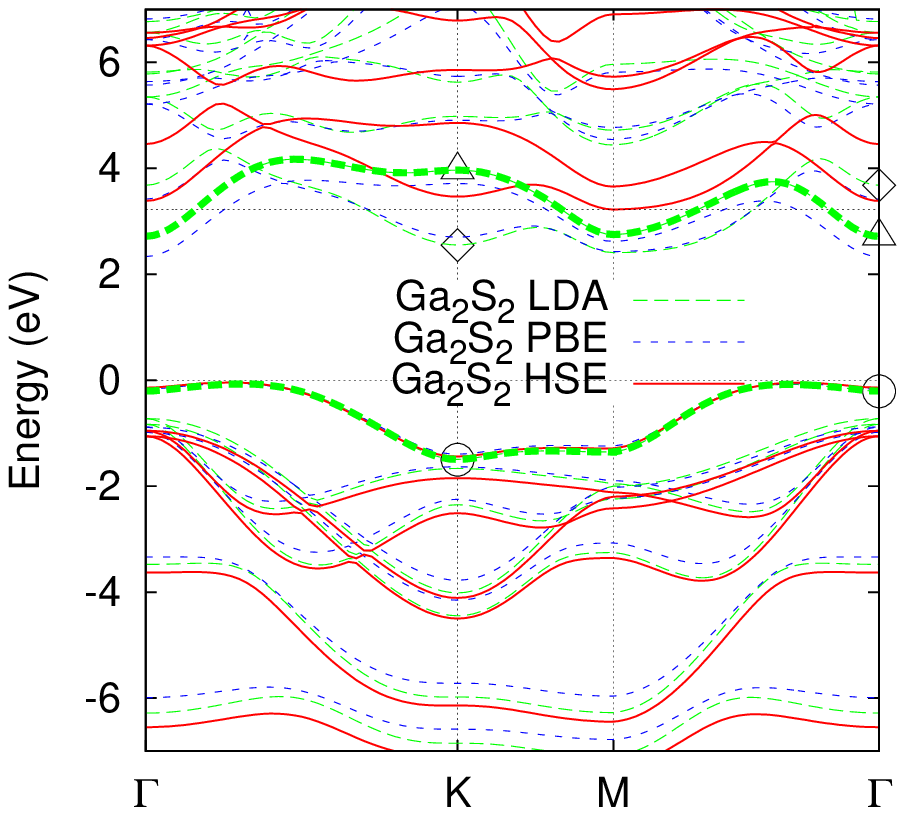}
\includegraphics[clip,scale=0.8]{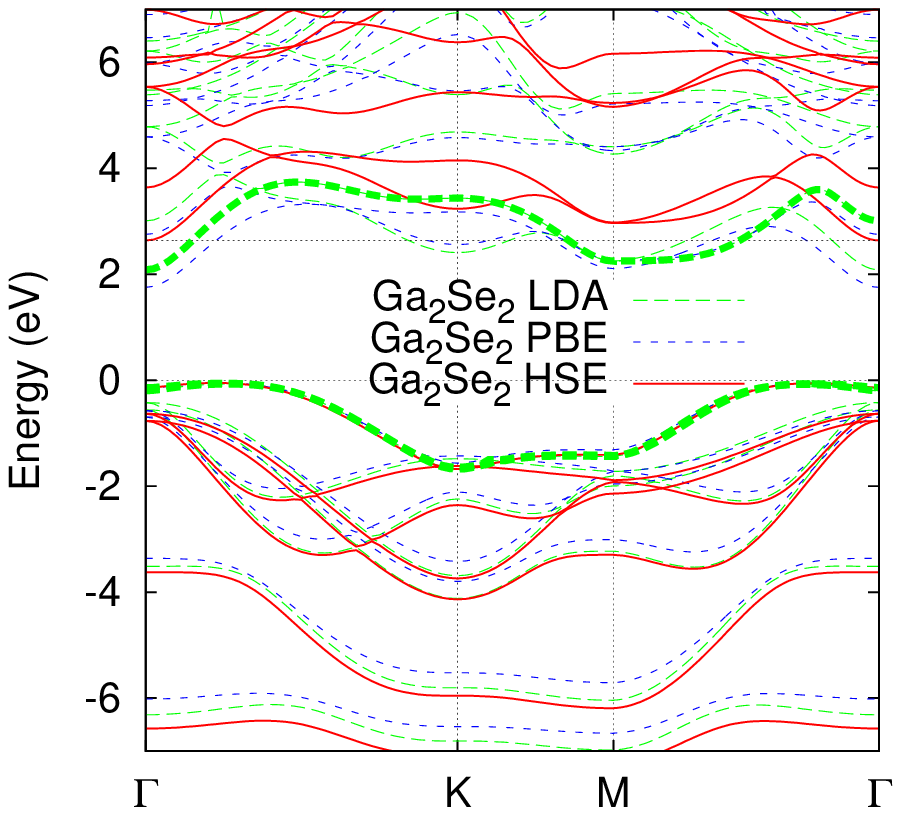}
\includegraphics[clip,scale=0.8]{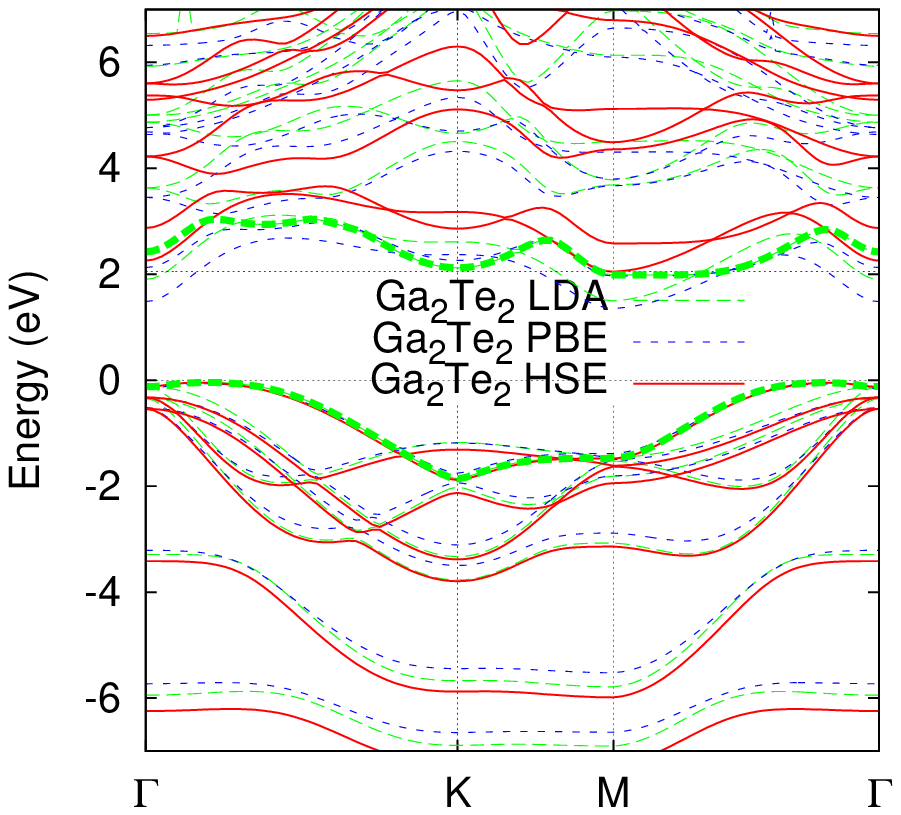}
\caption{(Color online) HSE06 band structures (solid red lines) for Ga$_2$S$_2$ (top), Ga$_2$Se$_2$ (middle),
and Ga$_2$Te$_2$ (bottom). The zero of energy is taken to be the Fermi level and the
bottom of the conduction band is marked with a horizontal line. For comparison, the semilocal
band structures are also shown. In the case of the LDA
the two symmetric bands that give rise to the main feature in the optical absorption spectrum are
marked in bold.
\label{fig:bands}
}
\end{center}
\end{figure}

Full geometry optimization was performed using both the LDA and PBE functionals. Table
\ref{table:parameters_summary} shows that the lattice constants increase with
the atomic number of the chalcogen atom, while the Ga-Ga bond lengths hardly change.
The bond lengths obtained with the PBE functional are systematically larger than those
optimized with the LDA, as expected \cite{FavotF_1999}.

\begin{table}
\caption{Structural parameters
(as defined in Fig.\ \ref{fig:structure}) of single-layer Ga$_2$X$_2$ (in~\AA)
from the DFT calculations.
\label{table:parameters_summary}}

\begin{tabular}{lcccccc}

\hline
\hline
X&$a^{\rm{LDA}}$&$d_{\rm{Ga}-\rm{Ga}}^{\rm{LDA}}$&$d_{\rm{X}-\rm{X}}^{\rm{LDA}}$&$a^{\rm{PBE}}$&$d_{\rm{Ga}-\rm{Ga}}^{\rm{PBE}}$&$d_{\rm{X}-\rm{X}}^{\rm{PBE}}$\\

\hline

S & $3.53$ & $2.41$ & $4.57$ &$3.63$&$2.47$&$4.64$\\

Se & $3.71$ & $2.40$ & $4.73$ &$3.82$&$2.47$&$4.82$\\

Te & $4.01$ & $2.40$ & $4.92$ &$4.14$&$2.47$&$5.02$\\

\hline
\hline

\end{tabular}
\end{table}

\begin{table}
\caption{
The band gaps of Ga$_2$X$_2$ according to the HSE06 functional
($\Delta_{HSE}$) and the spin-orbit splitting in the LDA
band structure at the K point in the
highest valence ($\Delta E_{SO}^{v,\textrm{K}}$) and lowest conduction ($\Delta E_{SO}^{c,\textrm{K}}$) bands.
\label{table:parameters_summary2b}}

\begin{tabular}{lccc}

\hline
\hline
X&$\Delta_{HSE}$&\quad $|\Delta E_{SO}^{v,\textrm{K}}|$ \quad \quad &$|\Delta E_{SO}^{c,\textrm{K}}|$\\

\hline

S & $3.28$ eV&$19$ meV&$21$ meV\\

Se &$2.69$ eV&$2$ meV&$36$ meV\\

Te &$2.10$ eV&$14$ meV&$102$ meV\\

\hline
\hline

\end{tabular}
\end{table}

\begin{table}
\caption{
The effective masses ($m^{\ast}/m_e$) of Ga$_2$X$_2$ at the high-symmetry points
in the conduction band according to the HSE06 functional
(in units of electron mass).
\label{table:parameters_summary2a}}

\begin{tabular}{lcccc}

\hline
\hline
X&$\Gamma^c$&K$^c$&M${^c}_{\rightarrow \Gamma}$&M${^c}_{\rightarrow \textrm{K}}$\\

\hline

S & $0.23$ & $0.62$&$1.52$&$0.29$\\

Se & $0.17$ & $0.56$&$3.03$&$0.23$\\

Te & $0.14$ & $0.47$&$0.43$&$0.19$\\

\hline
\hline

\end{tabular}
\end{table}

The calculated electronic band structures are summarized in Fig.\ \ref{fig:bands}. All three materials are indirect-gap semiconductors,
primarily due to the valence-band maximum (VBM) lying somewhat off the $\Gamma$ point.
It is possible to fit an inverted Mexican-hat polynomial to the valence-band dispersions $E_{VB}$
around the valence band maximum:

\begin{eqnarray}
\label{eq:one}
E_{VB} &=& \sum_{i=0}^{3}{E_{2i}k^{2i}} + {E^{\prime}_{6}k^{6}}\cos{6 \varphi},
\end{eqnarray}

\noindent where $k$ and $\varphi$ are the polar coordinates
of the wave vectors measured from the $\Gamma$ point in units of 1/\AA, $\varphi$ is measured from the $\Gamma-\textrm{K}$
line, and the energy is in units of eV\@. The coefficients are listed
in Table \ref{table:orbital_decomposition2}. This fit should provide a good starting point
for a simple analytical model of the valence band in these materials. Note however that this fit
is designed to describe the immediate vicinity of the VBM and the saddle point, and it no longer describes
the $\Gamma$ point correctly, overestimating the $\Gamma$ point valence band energy by
52 meV, 70 meV, and 32 meV in Ga$_2$S$_2$, Ga$_2$Se$_2$, and Ga$_2$Te$_2$, respectively.

\begin{figure}
\begin{center}
\includegraphics[clip,scale=0.7]{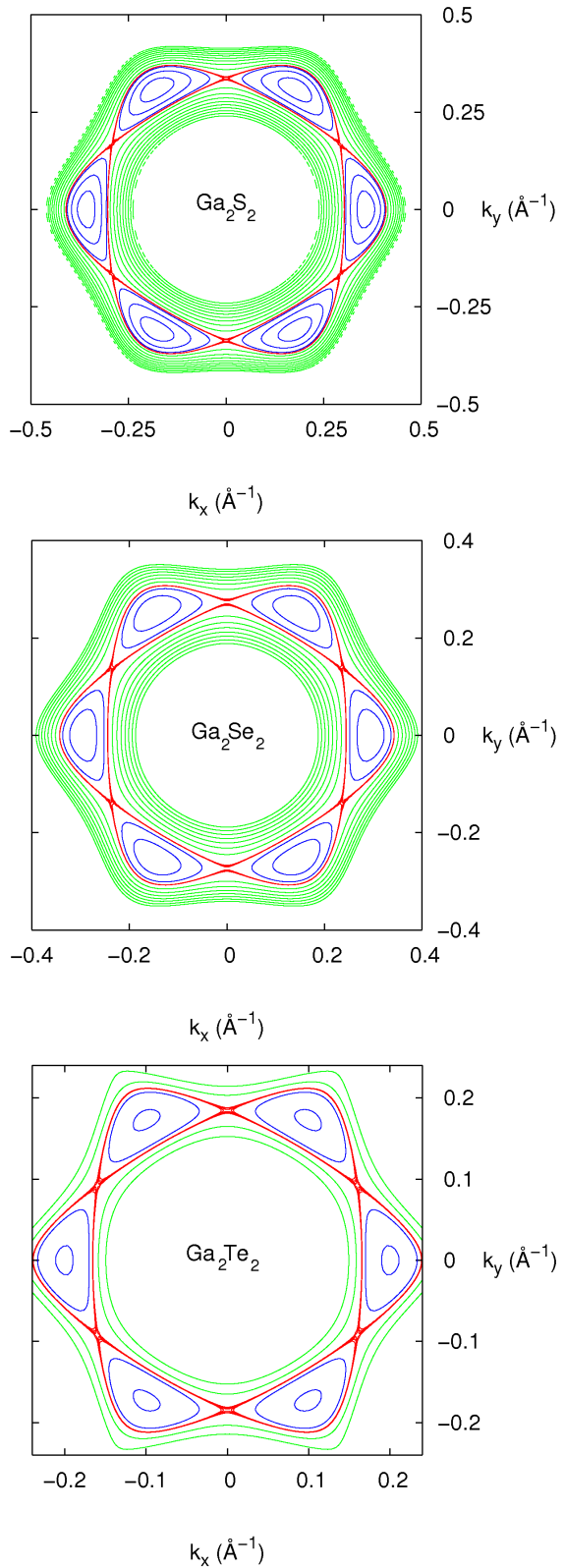}
\caption{(Color online) Energy contours of GaX in the valence band using the fitted formula of
Eq. (\ref{eq:one}). The separation of contours is 2 meV in all figures and the contour corresponding
to the energy of the saddle point along the $\Gamma-\textrm{M}$ line is highlighted.
\label{fig:contours}
}
\end{center}
\end{figure}

Further analysis of the valence band reveals a saddle point along the $\Gamma-\textrm{M}$ line,
illustrated in Fig. \ref{fig:contours}.
This saddle point gives rise to a Van Hove singularity in the density of states quite close to the
Fermi level, which is found to be at the valence band edge. Due to the presence of these
saddle points, hole-doping causes Ga$_2$X$_2$ to undergo a Lifshitz transition when the hole concentration
reaches the critical value where all states are depleted above the energy of the saddle point, since
this leads to a change in the topology of the Fermi surface. The carrier density where the Lifshitz
transition takes place in each material is listed in the last column of Table \ref{table:orbital_decomposition2}
and was obtained by integrating the DFT density of states from the saddle point to the valence band edge.

We find that the conduction-band minimum (CBM) is at the M point
in Ga$_2$S$_2$ and Ga$_2$Te$_2$, and at the
$\Gamma$ point in Ga$_2$Se$_2$. The HSE06 band gaps of Ga$_2$X$_2$
are summarized in Table \ref{table:parameters_summary2b}.
The HSE06 band gap is expected to underestimate the quasiparticle band gap by
no more than 10\ \% \cite{hse10percent}, and is known to be applicable
to two-dimensional materials as well \cite{hse10scuseria}.
Our finding that
the gap is indirect is in agreement with a recent DFT result obtained for single-layer Ga$_2$Se$_2$
\cite{rybkovskiy_dv_2006}, although density-functional
tight-binding calculations \cite{kohler_t_2004}
disagree with our results for Ga$_2$S$_2$\@. The effective masses at the high-symmetry points in the conduction band
are summarized in Table \ref{table:parameters_summary2a}. The effective mass is isotropic at $\Gamma$ and K, but not at the M point.

The semilocal band structures are also plotted in Fig.\ \ref{fig:bands} for comparison. The LDA and PBE functionals give very similar
results to the HSE06 functional up to the Fermi level, but above that significant discrepancies arise. This is
most notable in the case of Ga$_2$S$_2$, where the position of the CBM is ambiguous: the LDA predicts that
the CBM is at the M point, in agreement with HSE06, while the PBE functional puts it at the $\Gamma$ point.
In Ga$_2$Se$_2$ and Ga$_2$Te$_2$ both the LDA and the PBE functionals predict that the CBM is at the same
place as in the HSE06 calculation.

\begin{figure}
\begin{center}
\includegraphics[clip,scale=0.8]{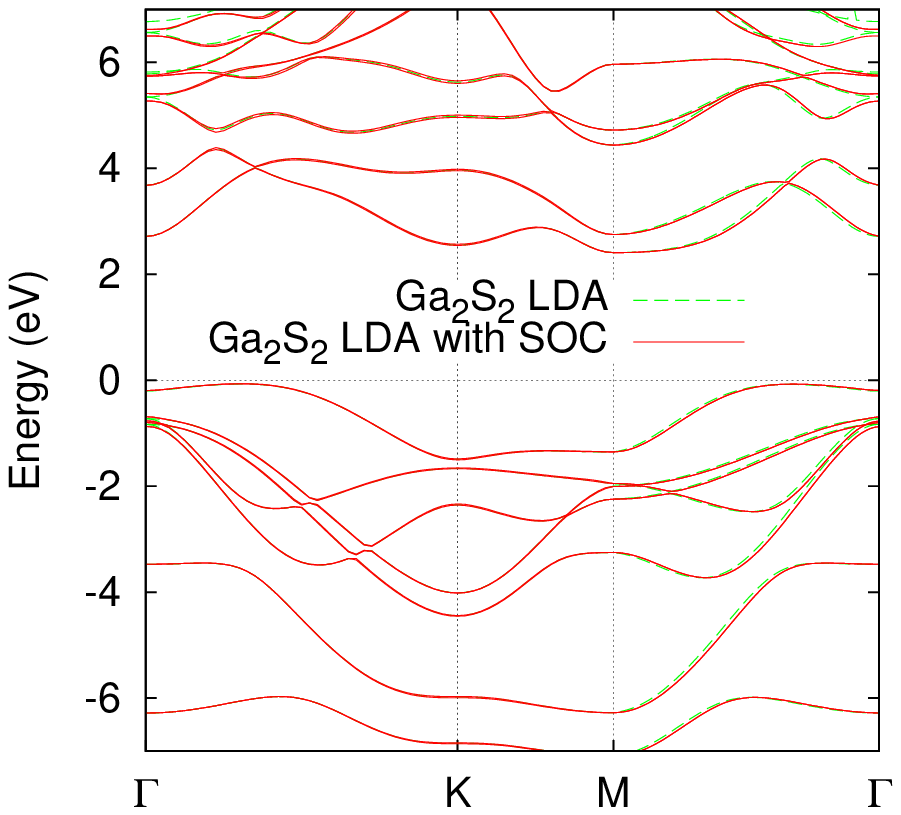}
\includegraphics[clip,scale=0.8]{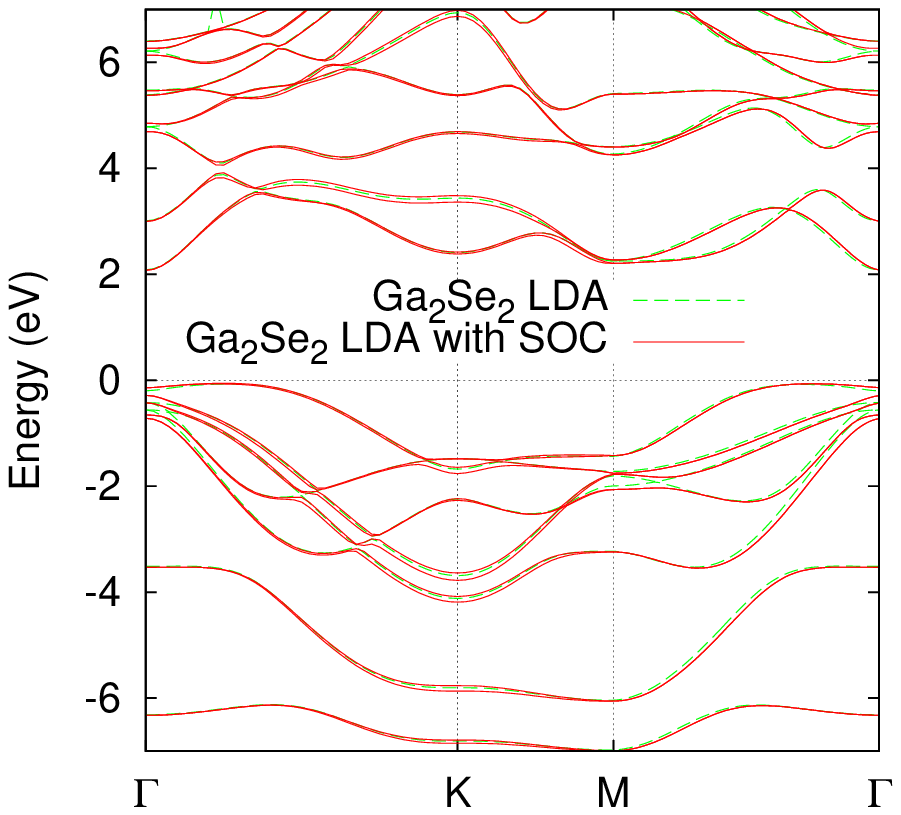}
\includegraphics[clip,scale=0.8]{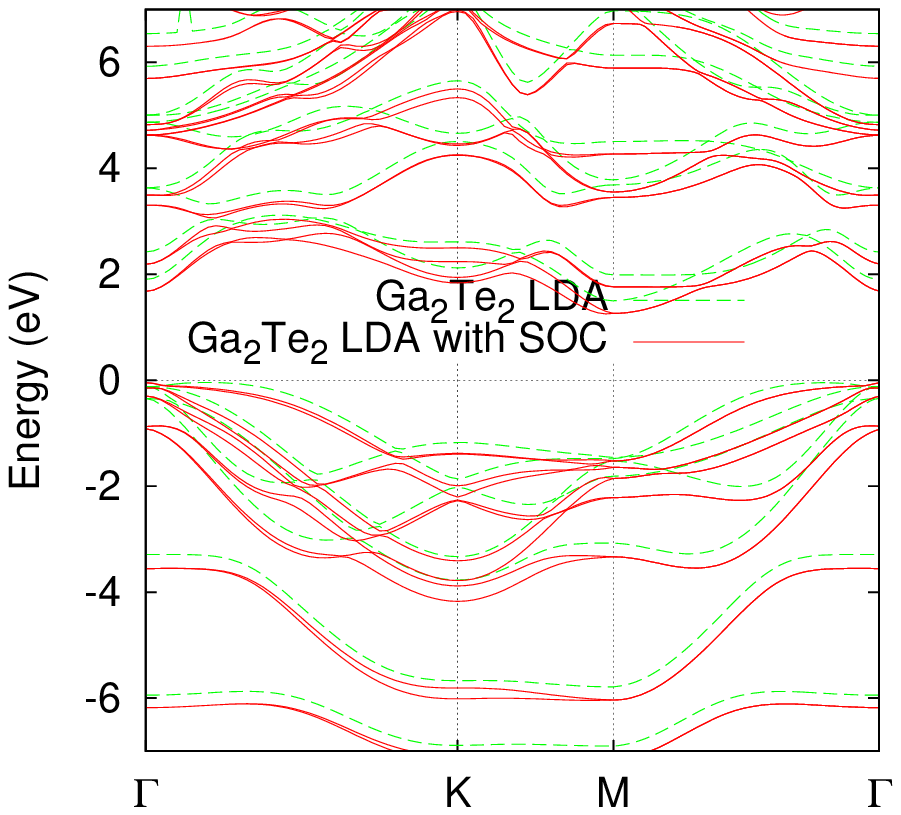}
\caption{(Color online) Comparison between the LDA band structures of Ga$_2$X$_2$ with
and without spin-orbit coupling (SOC).
\label{fig:bands2}
}
\end{center}
\end{figure}

In the case of the semilocal DFT calculations we also took spin-orbit coupling into account
using a relativistic DFT approach \cite{vasp}. As can be seen in Fig.\ \ref{fig:bands2} some
of the bands exhibit spin splitting, including the highest valence ($\Delta E_{SO}^{v,\textrm{K}}$) and
lowest conduction ($\Delta E_{SO}^{c,\textrm{K}}$) bands near the
K point, most significant in Ga$_2$Te$_2$ (see Table \ref{table:parameters_summary2b}). While we were unable
to calculate the spin-orbit splittings in HSE06 due to limited computational
resources, we expect that they will exhibit a similar magnitude to that found in the semilocal
band structures.

\begin{figure}
\begin{center}
\includegraphics[clip,scale=0.8]{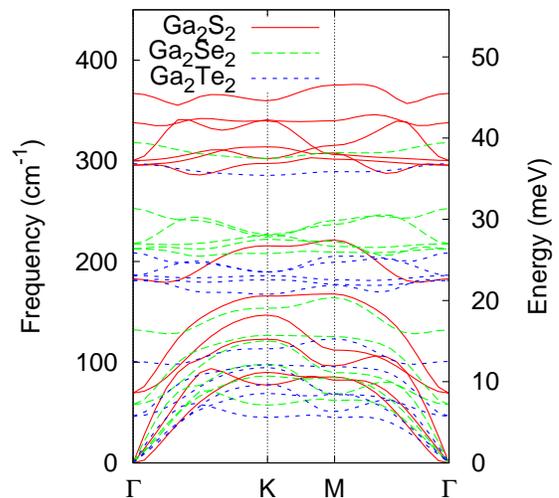}
\caption{(Color online) LDA phonon dispersion curves for Ga$_2$S$_2$, Ga$_2$Se$_2$, and Ga$_2$Te$_2$.
\label{fig:phonons}
}
\end{center}
\end{figure}

We have
calculated the phonon dispersions for Ga$_2$X$_2$ using the force-constant approach in up to
$4\times 4$ supercells.
The results are depicted in Fig.\ \ref{fig:phonons},
suggesting that isolated atomic crystals of gallium chalcogenides,
Ga$_2$X$_2$, are dynamically stable.

Table \ref{table:orbital_decomposition} shows an orbital composition analysis of the LDA bands
of Ga$_2$S$_2$  around the Fermi level. We have found that these bands are dominated by
$s$- and $p$-type orbitals. Furthermore, states in each band are either odd or even with respect to
$z\rightarrow -z$ symmetry (this information is obtained from the complex phases of the
orbital decomposition in \textsc{vasp}). Therefore, the interband absorption selection rules require that photons
polarized in the plane of the two-dimensional crystal are absorbed by transitions between bands
with the same $z\rightarrow -z$ symmetry of wave-functions (even$\rightarrow$even,
odd$\rightarrow$odd), and photons polarized along the $z$ axis cause transitions between bands
with opposite symmetry (even$\rightarrow$odd, odd$\rightarrow$even). Note that in the case of Ga$_2$Te$_2$
these numbers only provide a qualitative description of whether the orbitals are of mostly $s$ or
$p$ character due to the strong spin-orbit coupling.

\begin{table*}
\caption{Orbital analysis of the LDA bands around the Fermi level in Ga$_2$S$_2$ at the $\Gamma$ and
K points. The labels $\bigcirc$, $\triangle$, and $\diamondsuit$ correspond to the bands
labeled as such in Fig.\ \ref{fig:bands}. The data were
obtained by projecting the orbitals in the plane-wave basis set of \textsc{vasp}
onto spherical harmonics.
Dominant contributions were found to originate from $s$ and $p$
type orbitals; $+-$ subscripts refer to even ($+$) and
odd ($-$) states with respect to $z\rightarrow -z$ reflection.
\label{table:orbital_decomposition}}

\begin{tabular}{lccc}

\hline
\hline
Band&$\Gamma$&$$&K\\
\hline
$\bigcirc_{+}$&0.014$s^{\rm{Ga}}$+0.058${p_z}^{\rm{Ga}}$+0.004$s^{\rm{S}}$+0.189${p_z}^{\rm{S}}$&&0.058$s^{\rm{Ga}}$+0.169${p_z}^{\rm{Ga}}$+0.044${{p_{x}}{p_{y}}^{\rm{S}}}$\\
$\triangle_{-}$&0.142$s^{\rm{Ga}}$+0.008${p_z}^{\rm{Ga}}$+0.098$s^{\rm{S}}$+0.056${p_z}^{\rm{S}}$&&0.190$s^{\rm{Ga}}$+0.008${p_z}^{\rm{Ga}}$+0.051${{p_{x}}{p_{y}}^{\rm{S}}}$\\
$\diamondsuit_{+}$&0.075$s^{\rm{Ga}}$+0.107${p_z}^{\rm{Ga}}$+0.082$s^{\rm{S}}$+0.001${p_z}^{\rm{S}}$&&0.030${{p_{x}}{p_{y}}^{\rm{Ga}}}$+0.036${{p_{x}}{p_{y}}^{\rm{S}}}$\\
\hline
\hline

\end{tabular}

\end{table*}

\begin{table}
\caption{
Coefficients E$_{2i}$ (in units of eV\AA$^{2i}$) for the inverted Mexican-hat dispersion near the VBM in Eq. (\ref{eq:one}).
The root mean square of the residuals in Ga$_2$S$_2$, Ga$_2$Se$_2$, and Ga$_2$Te$_2$
was found to be 0.48, 0.68, and 0.67 meV respectively. The fitting parameters were obtained by fitting the formula of
Eq. (\ref{eq:one}) to the DFT data within the range of $0.28<|\textbf{k}|<0.42$,
$0.22<|\textbf{k}|<0.36$, and $0.12<|\textbf{k}|<0.26$ in Ga$_2$S$_2$, Ga$_2$Se$_2$, and Ga$_2$Te$_2$, respectively
(with \textbf{k} in units of \AA$^{-1}$). The zero of energy is set to
the valence-band maximum. The last column shows the critical hole concentration ($n_{X}$, in units of cm$^{-2}$) where
the Lifshitz transition takes place (see text).
\label{table:orbital_decomposition2}}
\begin{tabular}{lccccc|c}

\hline
\hline
X&E$_{0}$&E$_{2}$&E$_{4}$&E$_{6}$&E$^{\prime}_{6}$&$n_{X}$\\
\hline
S  & $-0.086$ &1.38  & $-6.8$  & 5.6 & 1.634&$7.96\cdot 10^{13}$\\
Se & $-0.059$ & 1.49 & $-11.4$ & 17  & 4.49 &$6.13\cdot 10^{13}$\\
Te & $-0.044$ & 2.50 & $-47$   & 240 & 30.1 &$3.54\cdot 10^{13}$\\
\hline
\hline

\end{tabular}

\end{table}

\begin{figure}
\begin{center}
\includegraphics[clip,scale=0.8]{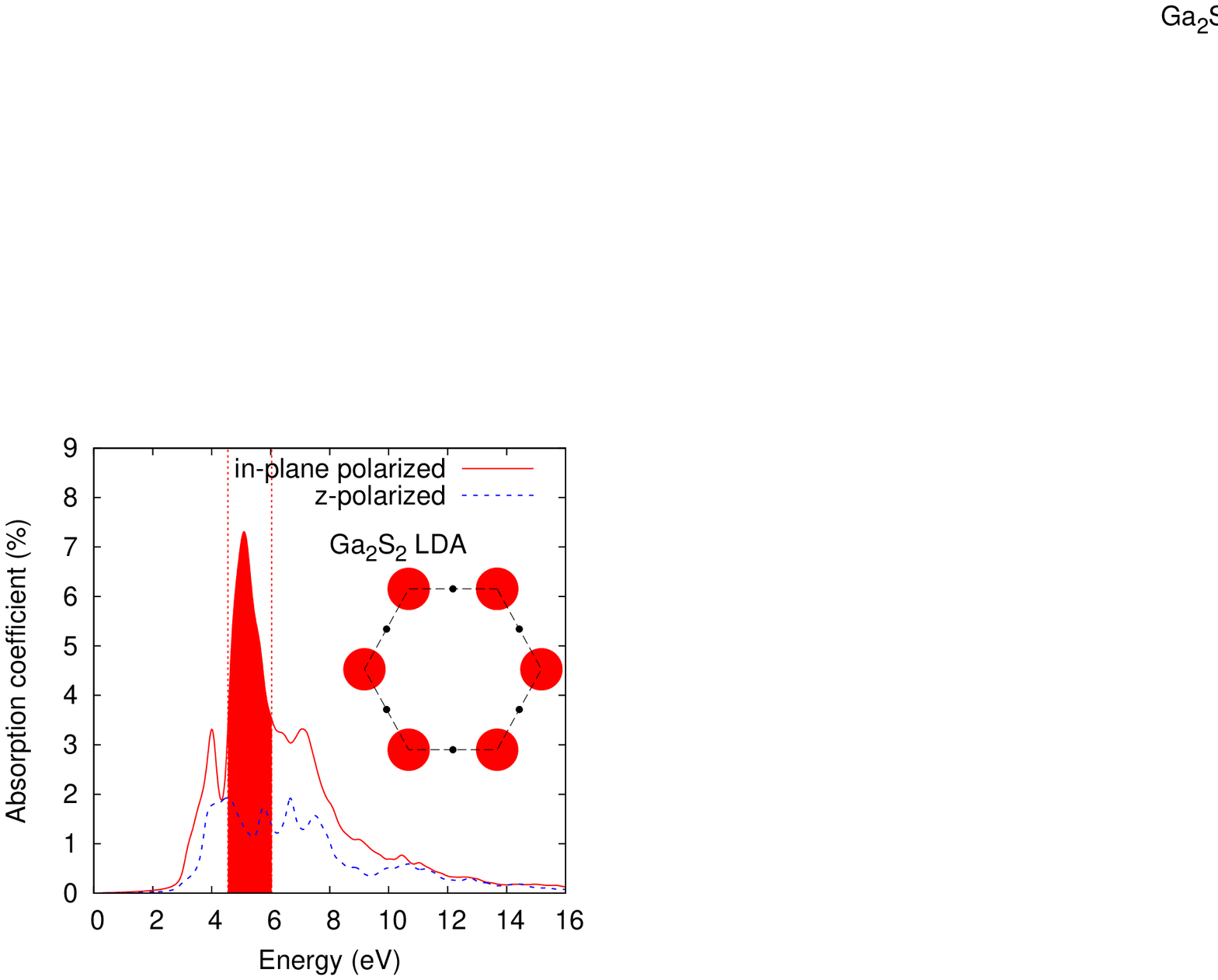}
\includegraphics[clip,scale=0.8]{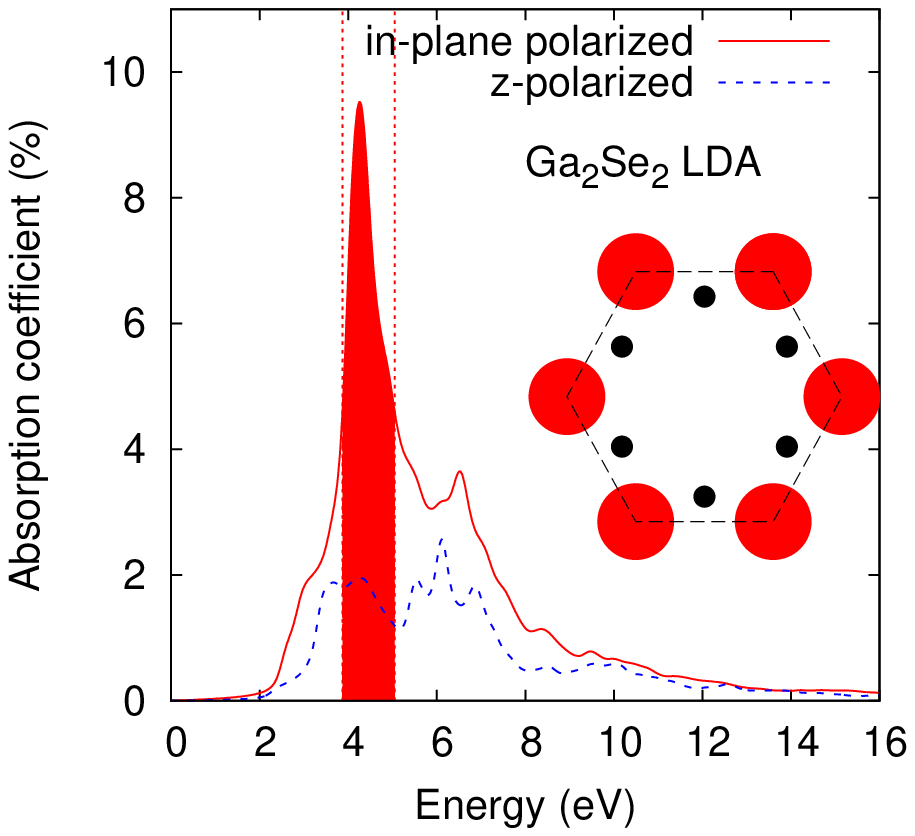}
\includegraphics[clip,scale=0.8]{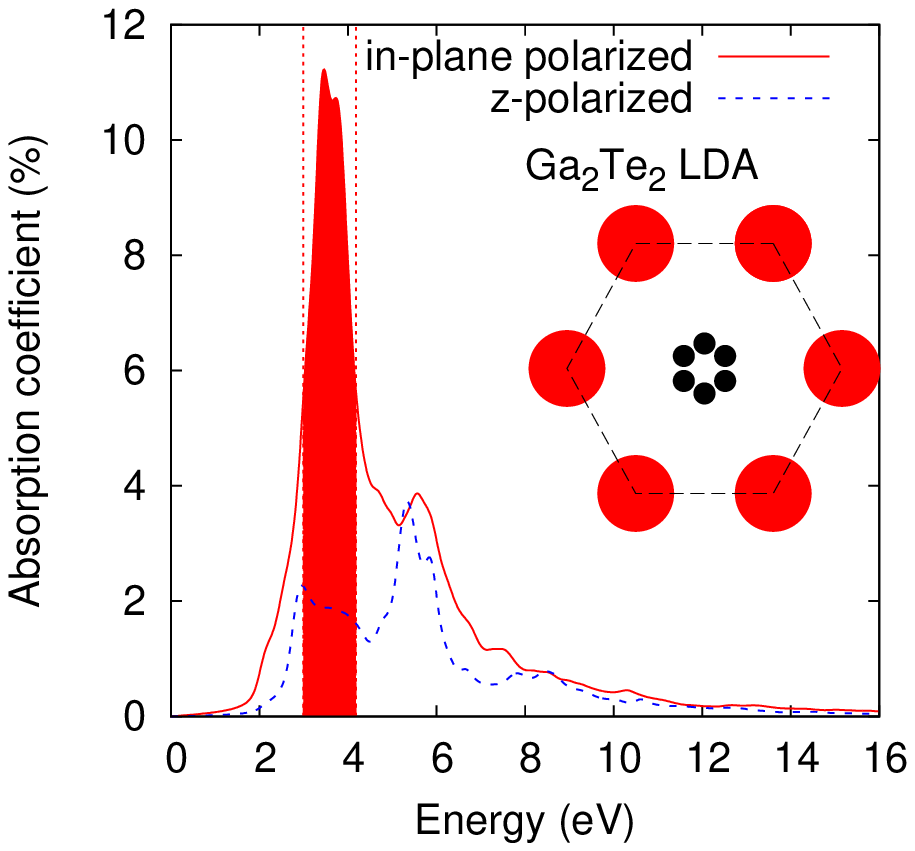}
\caption{(Color online) Absorption coefficient (the imaginary part of the dielectric
function $\varepsilon$) of various Ga$_2$X$_2$ two-dimensional crystals normalized to
absolute units after it was compared to Im$(\varepsilon)$ evaluated for graphene
in the range 0.8--1.5 eV, where monolayer graphene absorbs 2.3~\% of light (see text for details).
Vertical lines mark the full width at half maximum for the main peak.
The insets illustrate the relevant part of \textbf{k}-space for the transitions
between bands marked by circle and diamond in Fig.\ \ref{fig:bands} which give rise to the
main peak
(light red circle) and the low-energy
shoulder (black circle).
\label{fig:absorptions}
}
\end{center}
\end{figure}

The calculated optical absorption spectra (using semilocal DFT) are shown
in Fig.\ \ref{fig:absorptions}.
The intensities were normalized by using graphene as a benchmark. We calculated the absorption
of graphene using the same method in the 0.8--1.5 eV energy range where monolayer graphene
absorbs 2.3~\% of light, and used the calculated graphene absorption intensity to obtain a scaling
factor that scales the calculated intensities to the experimental value.
Since in the case
of Ga$_2$S$_2$ we found disagreement between the PBE and the HSE06 regarding the location of the CBM,
we restrict ourselves to the LDA in the following.
Note that the LDA results are only qualitatively accurate and should only be used for a comparative
study of the different Ga$_2$X$_2$ monolayers and for an order-of-magnitude estimate of the expected
peak positions. A better description would require a computationally much more expensive calculation
using the GW approximation and the Bethe-Salpeter equation for excitonic corrections \cite{bse_example}. 
In Fig.\ \ref{fig:absorptions} we show
the energy dependence of the absorption coefficients of various stoichiometric
Ga$_2$X$_2$ monolayers. The absorption spectrum starts with a shoulder
at low energies, which originates from the vicinity of
the M and $\Gamma$ points.
In all cases the in-plane absorption is much more significant and is
dominated by peaks at around 5 eV/4 eV/3.5 eV for X=S/Se/Te.
Analysis of the band structure reveals that the bulk of this spectrum comes from transitions
between the valence band and the $z\rightarrow -z$ even band above the conduction band
around the K point where the two bands are near-parallel (this is the only part in the band structure
where the energy difference between a filled and an empty band matches the energy of the main peak).
In that range, the absorption coefficients of Ga$_2$X$_2$ are comparable to and
even exceed that of monolayer and bilayer graphene; therefore we suggest that ultrathin films of
GaX biased in vertical tunneling transistors with graphene electrodes could be used as
an active element for detection of ultraviolet photons.

\section{Conclusion}

We have shown using first-principles density-functional-theory that two-dimensional
Ga$_2$X$_2$ (X=S, Se, and Te) crystals are stable indirect-band-gap semiconductors
with an unusual inverted Mexican-hat valence band. The presence of saddle points along
the $\Gamma-\textrm{M}$ line leads to a Lifshitz transition in the event of hole
doping for which we have calculated the critical carrier density. We have provided an analytical fit
of the valence-band edge and given a qualitative description of the optical absorption spectra,
which suggest that ultrathin films of
GaX biased in vertical tunneling transistors with graphene electrodes could be used as
an active element for detection of ultraviolet photons.

\begin{acknowledgments} We acknowledge financial support from
the EPSRC Science and Innovation Award, the ERC Advanced Grant ``Graphene and Beyond,''
the Royal Society Wolfson Merit Award, the Marie Curie project
CARBOTRON, and the EC STREP ``ConceptGraphene.''
\end{acknowledgments}

\end{document}